\documentstyle[twocolumn,prl,aps,epsfig]{revtex}

\begin{document}
\draft
\preprint{HEP/123-qed}
\title{Photonic quantum ring}
\author{J. C. Ahn, K. S. Kwak, B. H. Park, H. Y. Kang, J. Y. Kim,
and O'Dae Kwon}
\address{Department of Electronic and Electrical Engineering,
Pohang University of
Science and Technology, Pohang, 790-784, Korea.\\
E-mail: odkwon@postech.ac.kr}
\date{\today}
\maketitle
\begin{abstract}                
We report a quantum ring-like toroidal cavity naturally
formed in a vertical-cavity-like active microdisk plane
due to Rayleigh's band of whispering gallery modes.
The $\sqrt{T}$-dependent redshift and a square-law property
of microampere-range threshold currents down to 2 $\mu$A
are consistent with a photonic quantum wire view, due to
whispering gallery mode-induced dimensional reduction.
\end{abstract}

For the last several years, there have been intensive developments of
microdisk semiconductor lasers of whispering gallery(WG) modes for
low-power and high-density photonic array applications.
These efforts of earlier photo-pumped thumbtack-type WG lasers \cite{slusher}
have been evolving to photonic-wire lasers \cite{p-wire} and electro-pumped
thumbtack-type WG lasers \cite{levi,baba} of submilliampere threshold currents.
In this letter, we demonstrate a cylindrical
vertical-cavity surface-emitting laser(VCSEL)-like
diode that exhibits WG modes with unusual photonic
quantum ring(PQR) characteristics
such as $\mu$A-range threshold currents and spectral $\sqrt{T}$-dependence,
in addition to the usual VCSEL mode.
We will thus illustrate and analyze the two-threshold behavior of
successive lasings, first the PQR and then the VCSEL as well.
Indeed, the quantum wire behavior of the PQR is vividly demonstrated in the
$\sqrt{T}$-dependent spectral peak shift data.
We note that the quantum-ring-like WG modes are naturally formed in the
circumferential Rayleigh band region \cite{Rayleigh} of
the active microdisk of a regular multi-quantum-well
VCSEL-like structure but without any intentional and
real quantum ring patterning.

The PQR device differs from the previous WG
lasers \cite{slusher,p-wire,levi,baba,ury} in that the vertical confinement is
improved by the top and bottom DBR layers, and that stripe or segmented top
electrodes are used for vertical output coupling.
The metal-organic vapor-phase epitaxy-grown device
employs a one-$\lambda$ thick active layer with
four 80 {\AA} Al$_{0.11}$Ga$_{0.89}$As quantum wells(QWs),
whose details have been described elsewhere \cite{oecc97,spie,jjap}.
In the active disk plane, the PQR region is defined by Rayleigh's bandwidth,
$W_{Rayleigh} = (\phi/2)(1-n_{eff}/n)$, where the WG mode occurs,
which is well described in Ref.\ \onlinecite{ho}:
Rayleigh's annular band is defined by the active disk's
radius $R (= \phi/2$) for outer boundary and the inner
reflection point $r_{in} = R n_{M}/n$,
where $n$ is the refractive index of the active medium and
$n_{M}$ is the effective refractive index in azimuthal
direction($n_{M} \simeq n_{eff}$ \cite{slusher}).
\begin{figure}[tbh]
   \centering{\epsfig{figure=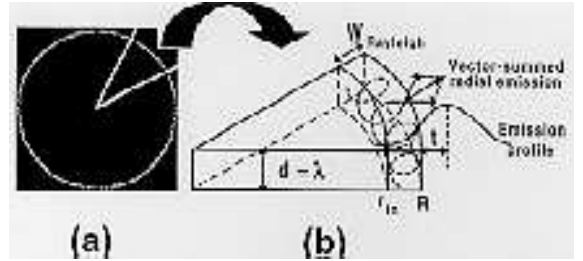, width=75mm}}
   \vspace{5mm}
   \caption{Calculated WG mode ring. (a) Annular Rayleigh band
   with $R = 7.5\ \mu$m, $W_{Rayleigh}=0.36\ \mu$m, and $M=188$.
   Azimuthal maxima region generates the WG ring while all other ripples
   are filtered out with -9 dB cutoff in this plot.
   (b) An exploded segment shows the effective toroidal structure
   of the PQR, where $R$ is the index-guided outer
   limit and $r_{in}$ is the inner reflection limit which may
   also be reinforced by gain guiding.
   The radial evanescent field results from forward and backward helical
   traveling waves, and its intensity peaks at a distance
   $t \leq 1\ \mu$m from the outer boundary.}
   \label{fig:wg-ring}
\end{figure}
The calculated profile of the annular WG emission based on Rayleigh's
Bessel function analysis is shown in Fig.\ \ref{fig:wg-ring}(a) for our
device with $R = 7.5\ \mu$m.
The azimuthal mode number is rather large here,
$M = 2\pi R n_{eff} / \lambda = 188$,
and hence the 376 azimuthal peaks are not well distinguishable,
unlike the intra-cavity axial VCSEL mode patterns
being easily resolved as shown below.
The exploded segment, Fig.\ \ref{fig:wg-ring}(b), shows cross-sectional
details of the one-$\lambda$ thick toroidal cavity of the PQR,
formed by the vertical DBRs, together with the in-plane annular
Rayleigh confinement, which then allows forward and backward helical
wave propagations of the intra-toroid WG modes, evanescently leading to
the extracavity emission in radial direction \cite{spie}.

\begin{figure}[tbh]
   \centering{\epsfig{figure=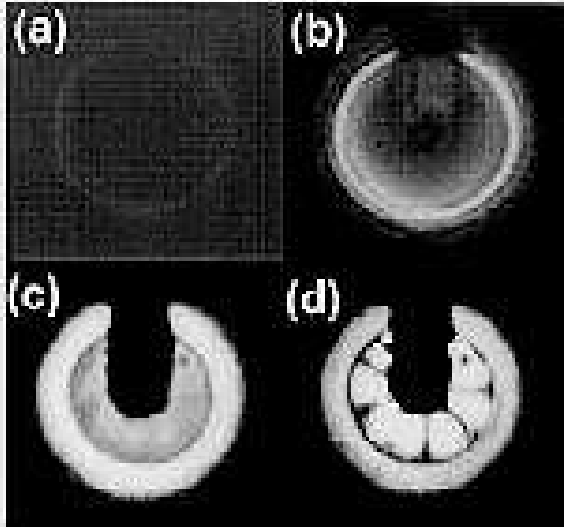, width=75mm}}
   \vspace{5mm}
   \caption{Microscope pictures of the PQR laser.
   The emission is partly blocked due to the stripe electrode.
   (a) I = 3 $\mu$A, where a faint evanescent mode image appears
   indicating the onset of the toroid transparency;
   (b) I = 12 $\mu$A, near the PQR threshold;
   (c) I = 11.5 mA, right below the VCSEL threshold;
   (d) I = 12 mA, at the VCSEL threshold.}
   \label{fig:photo}
\end{figure}

We now illustrate unusual two-threshold successive lasing behaviors with the
near-field micrographs of a 15-$\mu$m-diameter device as shown
in Fig.\ \ref{fig:photo}. The annular PQR emission patterns in
Figs.\ \ref{fig:photo}(a) and \ref{fig:photo}(b), exhibiting an onset of transparency
of $\sim$3 $\mu$A and a threshold of $\sim$12 $\mu$A respectively,
correspond to the conventional WG lasers, which is extracavity
evanescent field emission \cite{slusher}.
There occurs tremendous intensity buildup with increasing injection currents
so that neutral density filters were used in taking pictures of
Figs.\ \ref{fig:photo}(c) and \ref{fig:photo}(d).
Nevertheless, the ring-shape emission pattern and spectral behaviors remain
mostly unchanged. As the injection current increases above the
second(VCSEL) threshold level of $\sim$12 mA,
the emission of intra-VCSEL-cavity high-order transverse axial mode with
10-fold rotational symmetry
appears \cite{jjap}.
We also note that, after the VCSEL threshold,
the mode energies in the toroidal cavity split
into intracavity VCSEL mode and extracavity PQR WG modes, and that recent
studies of ultralow threshold VCSELs may have to consider this mode split loss.

We now mention some previous reports \cite{arbel,deppe} of observing WG modes
from VCSEL structures with distinguishable azimuthal peak patterns,
even though the observed WG mode patterns size-wise seem to be rather
intracavity-like. However, since the reports never discuss the temperature
dependence as described below, a direct comparison with our results is
impossible. Otherwise, perhaps due to submicron azimuthal mode spacings,
there has been no report of distinguishable WG patterns actually
observed \cite{slusher,baba}.

\begin{figure}[tbh]
   \centering{\epsfig{figure=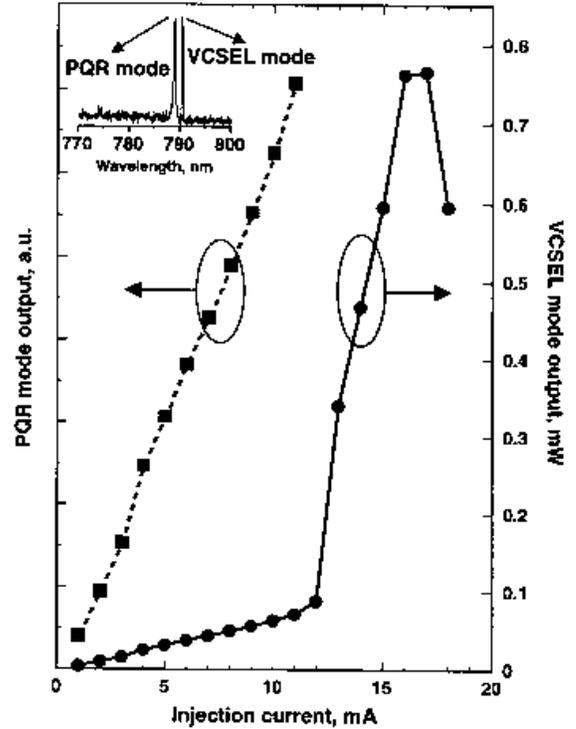, width=75mm}}
   \vspace{5mm}
   \caption{Light-current curves of the PQR and VCSEL modes.
   We note that data fluctuations occur in the low current region
   especially for the PQR spectral peak measurements.
   The microprobe for PQR measurements had a collection efficiency of
   $\sim 20 \%$ within a given radial segment. The inset shows
   typical spectra, at I = 5 mA for PQR and at 13 mA for VCSEL.}
   \label{fig:l-i}
\end{figure}

The room temperature CW light-current data for the PQR device
in Fig.\ \ref{fig:l-i} were collected by reading spectral peak intensities
that become resolvable for I $> 400\ \mu$A with a microprobe made
of a tapered single mode fiber and a spectrum analyzer (HP model 70951A).
After the VCSEL threshold, the portion of curve for the PQR emission was not
separable from that of the VCSEL, because of data overlapping due to
a monitoring microscope placed right above the device.
The near-vertical PQR emission, with an off-normal
probe angle of $\sim 5^{\circ}$, had a spectral linewidth of $\sim$0.5 nm as
compared with the VCSEL linewidth of $\sim$0.2 nm.
The broadened PQR linewidth due to peripheral surface roughness
will be discussed later.

\begin{figure}[tbh]
   \centering{\epsfig{figure=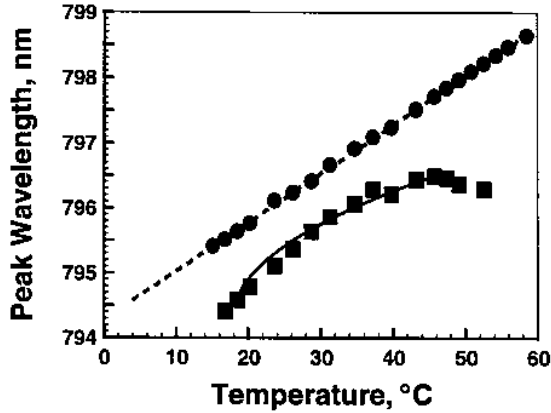, width=75mm}}
   \vspace{5mm}
   \caption{Temperature dependent spectra measured with a constant injection
   current, I = 12 mA. Circles represent spectral
   data for the VCSEL, while squares are for the PQR, indicating
   a $\sqrt{T}$-dependence.}
   \label{fig:temp}
\end{figure}

To distinguish the PQR WG mode from the VCSEL mode,
we present the wavelength shift data for both as a function of
the device temperature in Fig.\ \ref{fig:temp} \cite{tem-detail}.
It is notable that the spectral shift of the PQR mode
shows a distinct $\sqrt{T}$-dependence,
while the VCSEL mode wavelength increases linearly
at a rate of 0.07 nm/$^\circ$C. The best fits to both curves
in Fig.\ \ref{fig:temp} are given as
$\lambda_{PQR} = 0.42 \sqrt{T-18} + 794.3$ [nm] and
$\lambda_{VCSEL} = 0.07 T + 794.3$ [nm], respectively.
One of the results in Ref.\ \onlinecite{taylor} shows linear $T$-dependence
for both the index-dominated spectral change
for short resonant cavities like the VCSEL and the
gain-dominated spectral change for long cavities like
the edge-emitting diode. The above measured temperature
coefficient of 0.07 nm/$^\circ$C for the intracavity VCSEL mode
is consistent with the index-dominated mechanism. On the
other hand, the 3-dimensional Rayleigh toroid for the
PQR emission is to be classified as a long cavity confining
the helical wave propagation along the circumference of the
active disk, and hence the PQR emission
ought to be gain-dominated.
The observed spectral $\sqrt{T}$-dependence then implies a
gain-induced lasing, and is best explained by the
quantum wire assumption, i.e., $\sqrt{T}$-dependent
transparency condition \cite{yariv}.

We now pay special attention to the $\sqrt{T}$-dependent
characteristics of the PQR because the PQRs would occupy the peripheral region of
the four QW active disk, which should have a uniform carrier
distribution. However, the observed two-threshold
emission behavior indicates a significantly lowered transparency
level for the PQR region as seen in Fig.\ \ref{fig:photo}(a),
compared with that of the central region of the QW plane. A
possible reason for this is the Rayleigh confinement
mechanism leading to the 3-dimensional toroidal cavity. This peculiar
confinement appears to limit the number of guided resonant modes
drastically, which in turn reduces the amount of energy required
for the overall PQR mode excitation. Moreover, the above situation
may involve a possible universal mechanism of dimensional reduction \cite{comment},
and is also effectively
analogous to the enhancement of the spontaneous emission factor
$\beta$, which is the key interest of the various microcavity
structures suggested such as microdisk, microdroplet, and photonic
bandgap structures \cite{microcavity}.
We also note that wavelength shifts of the PQR mode
level off in the region of $T \geq 40^\circ$C,
which might solve the nagging problem of the spectral
wandering due to local device heating associated with
typical high-density laser arrays.

\begin{figure}[tbh]
 \centering{\epsfig{figure=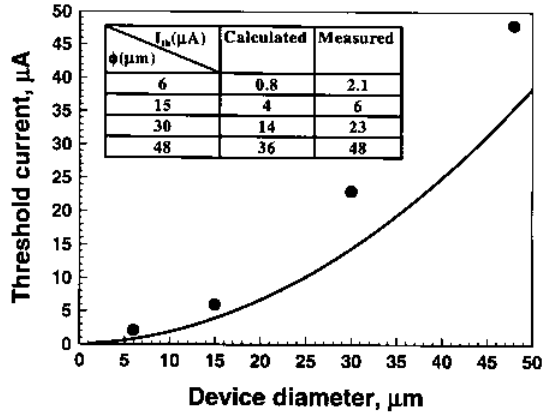, width=75mm}}
 \vspace{5mm}
 \caption{Size-dependent PQR threshold currents.
 The solid curve is from Eq.\ (\ref{eq:th-cur}) and the circles are from PQR threshold
 currents measured.}
 \label{fig:cal-th}
\end{figure}

The Rayleigh toroid, associated with the PQR emission as described
above, is to confine the WG modal manifolds of the
above-mentioned helically twisted traveling waves.
We thus propose a concept of 2-dimensional pseudo-quantum rings for
providing carriers needed for the excitation of the WG modal manifolds,
which would take place well below the transparency of the whole QW
plane \cite{hollow}.
In order to estimate the threshold current of the WG emission,
we now assume the toroidal structure of our device as a concentric array of
PQRs whose lateral characteristic unit length is defined by half the
wavelength($\lambda_{PQR}$/2), being much larger than de Broglie's wavelength,
to reflect the nature of paraxial
lightwave characterized by transverse peaks with a half-cycle interval.
For such a pseudo-quantum wire, assumed
to provide the carriers needed for the individual PQR, the transparency
carrier density is calculated by using a formula \cite{yariv},
$N_{tr} = (\sqrt{2m_{C}}/\pi \hbar) \times 1.072 \sqrt{k_{B}(T+273)}$.
The overall transparency current, $I_{tr}^{Rayleigh}$, can now be
calculated by multiplying
the transparency current, $I_{tr}^{PQR} = N_{tr} \pi \phi e / \eta \tau $,
for a single quantum-wire ring with the number($\chi$) of concentric rings
within the active Rayleigh WG band so that
$I_{tr}^{Rayleigh} \equiv \chi I_{tr}^{PQR}
= W_{Rayleigh}/(\lambda_{PQR}/2n_{eff}) N_{tr} \pi \phi e/(\eta \tau)$,
where $\phi$ is the device diameter,
$\eta$ the quantum efficiency, and $\tau$ the carrier lifetime.
The additional terms \cite{yariv} due to the intrinsic loss
$\alpha_{i}$ = 5 cm$^{-1}$ \cite{zory} and the mirror loss are then included in
the final expression for the threshold current:
\begin{eqnarray}
I_{th} & = & I_{tr}^{Rayleigh} + I_{i} + I_{mirror} \nonumber \\
       & = & (\frac{\pi n_{eff} e N_{tr} w}{\lambda_{PQR} \tau}) \phi^{2}
               + (\frac{\pi d e \alpha_{i} w}{2g_{1D}^{'} \tau}) \phi^{2} \nonumber \\
       & + & (\frac{d e w \ln R^{-1}}{2g_{1D}^{'} \tau}) \phi,
\label{eq:th-cur}
\end{eqnarray}
where $w \equiv 1 - n_{eff}/n = W_{Rayleigh}/R$ is a normalized Rayleigh
bandwidth, $d$ is the active region thickness,
and $g_{1D}^{'} = 8 \times 10^{-16}$ cm$^2$ \cite{zory}
is the differential gain coefficient.
This formula indicates a square-law property, $I_{th} \propto \phi^2$,
excluding the relatively small contribution from the DBR mirror loss.
A comparison between experiment and theory is shown in Fig.\ \ref{fig:cal-th},
and it indeed suggests an excellent agreement
except some systematic discrepancies presumably
due to substantial scattering losses
associated with microscopically rough peripheral structures \cite{spie}.
As the device diameter decreases, the Rayleigh bandwidth $W_{Rayleigh}$
also decreases, and in turn the scattering loss
due to the peripheral surface roughness
is now more serious and thus becomes the dominant loss factor.
This loss affects the observed threshold currents: for instance,
$I_{th} = 2\ \mu$A, larger than the calculated value
$I_{th} = 0.8\ \mu$A for $\phi = 6 \mu$m,
as tabulated in Fig.\ \ref{fig:cal-th}.
The peripheral roughness is also related to the broad
linewidth of the PQR mode, 0.5 nm, which is relatively large
compared with the VCSEL linewidth.
Refined epitaxy, lithography and etching
processes \cite{jjap} for smoother cylindrical surface,
minimizing the scattering loss \cite{slusher,baba}, will further narrow
the linewidth of the PQR emission like that of VCSELs.

In summary, we have observed ultralow threshold currents
of $\sim \mu$A range, and two-threshold successive lasings
of the PQR and VCSEL modes, from a toroidal cavity naturally formed
in a VCSEL-like microcavity structure. We have also shown the
$\sqrt{T}$-dependence of the spectral peak shift and a square-law
behavior of threshold currents, consistent with the PQR analysis.
Further work for lowered scattering loss is also under way for
achieving linewidth-narrowed, submicro-ampere laser diodes so that
the high-density integration of photonics to electronics become a
reality.

\acknowledgments
The authors thank Drs. R. E. Slusher, D. A. B. Miller,
and L. M. F. Chirovsky for helpful discussions, and
H. Han, H. K. Shin, and B. S. Yoo for technical assistance.
This work was supported by KT, Samsung Co., and KOSEF and ADD through KAIST.


\begin{references}
\bibitem{tem-detail}
Details of temperature experiments will be published elsewhere.
\bibitem{comment}
For an example in 2 dimension of the QW plane
with a uniform carrier distribution,
we may have a dynamically-generated carrier
concentration hump, a quantum-wire-like one-dimensional nature,
along the peripheral boundary. This is analogous to the soaring
water head at the moment of a collision with the wall of a pier
as the waves rush and collapse.
\bibitem{hollow}
We have recently found that hollow cylindrical cavities show
similar PQR threshold characteristics as well. This aspect will be
reported elsewhere.
\end{references}
\end{document}